\documentclass[12pt]{birkjour}
\usepackage{wrapfig}
\usepackage{amsmath,amsfonts,graphicx}
\usepackage{cite} 
\usepackage{hyperref}
\usepackage{epstopdf}

\begin{document}

\title[Exact solution of a generalized two-sites Bose-Hubbard model]{Exact solution of a generalized two-sites Bose-Hubbard model}

\author[Gilberto N. Santos Filho]{Gilberto N. Santos Filho}

\address{Centro Brasileiro de Pesquisas F\'{\i}sicas - CBPF \\
Rua Dr. Xavier Sigaud, 150, Urca, Rio de Janeiro - RJ - Brazil.}

\email{gfilho@cbpf.br}

\begin{abstract}
I introduce a new parametrization of a bosonic Lax operator for the algebraic Bethe ansatz  method with the $gl(2)$-invariant $R$-matrix and  use it to present the exact solution of a generalized two-sites Bose-Hubbard model with asymmetric tunnelling.  In the no interaction limit  I show that the Bethe ansatz equations can be written as a $S^{N-1}$ sphere, where $N$ is the total number of atoms in the condensate. 
\end{abstract}

\thanks{The author acknowledge Capes/FAPERJ (Coordena\c{c}\~ao de Aperfei\c{c}oamento de Pessoal de N\'{\i}vel Superior/Funda\c{c}\~ao de Amparo \`a Pesquisa do Estado do Rio de Janeiro) for financial support.}

\maketitle



\section{Introduction}

The first experimental verification of the Bose-Einstein condensation (BEC) \cite{ak,anderson,wwcch} occurred more then seven decades after its theoretical prediction \cite{bose,eins}, and a great deal of progress has been in the theoretical and experimental study of this many body physical phenomenon \cite{Dalfovo,l01,cw,Donley,Bloch,Carusotto}. In this direction the algebraic Bethe ansatz method has been used to solve and study models that may describe BEC \cite{GSantos11a,GSantos11b,GSantos13}.  The quantum phase transitions and classical analysis of some of these models have been studied in \cite{GSantos06a,GSantos09,GSantos10}. We are considering here a generalized issue of the two-sites Bose-Hubbard model, also known as the 
{\it canonical Josephson Hamiltonian} \cite{l01}, that has been an useful model in 
understanding tunnelling phenomena using two BEC \cite{albiez,milb,hines2,ours,our,GSantos06b,hines}.  The model that we will study is more general that the model \cite{l01,jlreview} in the sense that we introduce the on-well energies and asymmetric tunnelling. Here we will discuss its integrability and exact solution. The generalized model is described by the Hamiltonian 
\begin{eqnarray}
\hat{H} &=& \sum_{i,j=1}^2 K_{ij} \hat{N}_i\hat{N}_j - \sum_{i=1}^2(U_i - \mu_i) \hat{N}_i -   \sum_{i\neq j}^2 \Omega_{ij} \hat{a}_i^\dagger \hat{a}_j, 
\label{ham} 
\end{eqnarray}
\noindent where, $\hat{a}_i^\dagger (\hat{a}_i)$, denote the single-particle creation (annihilation) operators  and $\hat{N}_i = \hat{a}_i^\dagger \hat{a}_i$ are the corresponding  boson number operators in each condensate. The boson operator total number of particles, $\hat{N} = \hat{N}_1+\hat{N}_2$,  is a conserved quantity, $[\hat{H},\hat{N}]=0$. The couplings $K_{ij}$, with $K_{ij} = K_{ji} \; (i \neq j)$, provides the interaction strength between the  bosons and they are proportional to the $s$-wave scattering length,  $\Omega_{ij}$ are the   amplitude of tunnelling,  $\mu_i$ are the external potentials and $U_i = K_{ii} - \kappa_i$ are the on-well energies per particle,  with $\kappa_i$ the kinetic energies in each condensate.

The Hamiltonian (\ref{ham}) is  invariant under the discrete $\mathbb{Z}_2$ mirror transformation, $\hat{a}_j \rightarrow -\hat{a}_j$, and under the global  $U(1)$ gauge transformation, $\hat{a}_j \rightarrow e^{i\alpha}\hat{a}_j$, where $\alpha$ is an arbitrary $c$-number and, $\hat{a}^{\dagger}_j\rightarrow e^{-i\alpha}\hat{a}^{\dagger}_j,\;\;j=1,2$. For $\alpha = \pi$ we get again the $\mathbb{Z}_2$ symmetry.

 For the particular choice of the couplings parameters we can get some Hamiltonians, as for example  by the choices  $K_{ii} =  \frac{K}{8}$, $ K_{12} = K_{21} =  -\frac{K}{8}$, $\mu_1=-\mu_2 = \mu $, $U_i=0$, and $\Omega_{12} = \Omega_{21} = \frac{\mathcal{E}_{J}}{2}$ we get the canonical Josephson Hamiltonian studied in    \cite{l01}.  The  case with  $K_{12} = K_{21} = 0$, $K_{ii} =  U_i = U$, $\mu_1=-\mu_2 = \mu$, and $\Omega_{12} = \Omega_{21} = t$ was used  to study the interplay between disorder and interaction \cite{sarma}.  For these models we have symmetric tunnelling if $\Delta\mu = 0$ and when we turn on $\Delta\mu$ we break the symmetry. For the symmetric case we also can put $\mu_1=\mu_2=\mu$ and change the deep of both wells at the same time.  In the antisymmetric case $U_1 - \mu_1 \neq U_2 - \mu_2$ we have asymmetric tunnelling with the bias of one well  increasing the on-well energy. In this case it is called a tilted two-wells potential \cite{Dounas} and an experimental set up was made to study the distillation of a Bose-Einstein condensate, providing a model system for metastability in condensates, a test for quantum kinetic theories of condensate formation \cite{Shin} and atomtronic devices \cite{Seaman,Labouvie}. The on-well energies is  determined by the internal states of the atoms in the condensates and/or by the kinetic (thermal) energy of the atoms.

\begin{figure}[hb]
\begin{center}
\includegraphics[scale=0.2]{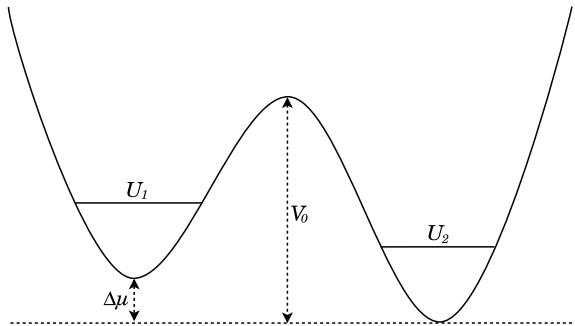}
\caption{Two-wells potential showing the asymmetric tunnelling for the case  $U_1 = U_2$, $\Delta\mu \neq 0$ and barrier height $V_0$.}
\label{tw}
\end{center}
\end{figure}

 In the Fig. \ref{tw} we represent the two BEC by a two-wells potential for the case  $U_1 = U_2$ and $\Delta\mu \neq 0$, with asymmetric tunnelling $\Omega_{12} \neq \Omega_{12}$. The tunnelling amplitudes $\Omega_{ij}$ are related to the barrier height $V_0$ \cite{l02}.
  
\section{The algebraic Bethe ansatz method}

The spectrum of the Hamiltonian (\ref{ham}), with $U_i = 0$ and symmetric tunnelling,  has been appeared in different papers \cite{jlletter,jlreview,jlsigma,Angela2,Angela,GSantos06b} and the Bethe states in \cite{Bethe-states}.  To be complete I will shortly describe the algebraic Bethe ansatz method \cite{jlreview,Angela} and present the exact solution  for the general case of $U_i \neq 0$ and asymmetric tunnelling $\Omega_{12} \neq \Omega_{12}$. We begin with the $gl(2)$-invariant $R$-matrix, depending on the spectral parameter $u$,

\begin{equation}
R(u)= \left( \begin{array}{cccc}
1 & 0 & 0 & 0\\
0 & b(u) & c(u) & 0\\
0 & c(u) & b(u) & 0\\
0 & 0 & 0 & 1\end{array}\right),
\end{equation}
\noindent with $b(u)=u/(u+\eta)$, $c(u)=\eta/(u+\eta)$ and $b(u) + c(u) = 1$. Above,
$\eta$ is an arbitrary parameter, to be chosen later.

It is easy to check that $R(u)$ satisfies the Yang-Baxter equation

\begin{equation}
R_{12}(u-v)R_{13}(u)R_{23}(v) = R_{23}(v)R_{13}(u)R_{12}(u-v),
\end{equation}

\noindent where $R_{jk}(u)$ denotes the matrix acting non-trivially
on the $j$-th and the $k$-th spaces and as the identity on the remaining
space.

Next we define the monodromy matrix  $\hat{T}(u)$,

\begin{equation}
\hat{T}(u)= \left( \begin{array}{cc}
 \hat{A}(u) & \hat{B}(u)\\
 \hat{C}(u) & \hat{D}(u)\end{array}\right),\label{monod}
\end{equation}
\noindent such that the Yang-Baxter algebra is satisfied

\begin{equation}
R_{12}(u-v)\hat{T}_{1}(u)\hat{T}_{2}(v)=\hat{T}_{2}(v)\hat{T}_{1}(u)R_{12}(u-v).\label{RTT}
\end{equation}
\noindent In what follows we will choose a realization for the monodromy matrix $\pi(\hat{T}(u))=\hat{L}(u)$  
to obtain a solution for the two-sites BEC model (\ref{ham}). 
In this construction, the Lax operator $\hat{L}(u)$  have to satisfy the algebra

\begin{equation}
R_{12}(u-v)\hat{L}_{1}(u)\hat{L}_{2}(v)= \hat{L}_{2}(v)\hat{L}_{1}(u)R_{12}(u-v),
\label{RLL}
\end{equation}
\noindent where we use the notation,
\begin{equation}
 \hat{L}_1 = \hat{L}(u) \otimes \hat{I} \;\;\; \mbox{and} \;\;\; \hat{L}_2 = \hat{I} \otimes \hat{L}(u).
\end{equation}

Then, defining the transfer matrix, as usual, through

\begin{equation}
\hat{t}(u)= \mbox{Tr} \;\pi(\hat{T}(u)) = \pi(\hat{A}(u) + \hat{D}(u)),
\label{trTu}
\end{equation}
\noindent it follows from (\ref{RTT}) that the transfer matrix commutes for
different values of the spectral parameter; i. e.,

\begin{equation}
[\hat{t}(u),\hat{t}(v)]=0, \;\;\;\;\;\;\; \forall \;u,\;v.
\end{equation}
\noindent Consequently, the models derived from this transfer matrix will be integrable. Another consequence is that the coefficients $\hat{\mathcal{C}}_k$ in the transfer matrix $\hat{t}(u)$,

\begin{equation}
\hat{t}(u) = \sum_{k} \hat{\mathcal{C}}_k u^k,
\label{poli-u}
\end{equation}
\noindent are conserved quantities or simply $c$-numbers, with

\begin{equation}
[\hat{\mathcal{C}}_j,\hat{\mathcal{C}}_k] = 0, \;\;\;\;\;\;\; \forall \;j,\;k.
\end{equation}

If the transfer matrix $\hat{t}(u)$ is a polynomial function in $u$, with $k \geq 0$, it is easy to see that,

\begin{equation}
\hat{\mathcal{C}}_0 = \hat{t}(0) \;\;\; \mbox{and} \;\;\; \hat{\mathcal{C}}_k = \frac{1}{k!}\left.\frac{d^k\hat{t}(u)}{du^k}\right|_{u=0}. 
\label{C14b}
\end{equation}

We are using a new solution of the equation (\ref{RLL}), a new parametrization of a well known \cite{GSantos06b,Angela} Lax operator,

\begin{equation}
\hat{L}_{i}(u)= \left(
\begin{array}{cc}
\lambda_i(u \hat{I} + \eta \hat{N}_i) & \alpha_i\hat{a}_i \\
\beta_i\hat{a}_i^{\dagger} & \alpha_i\beta_i\gamma_i\eta^{-1}\hat{I}
\end{array}
\right)\;\;\;\;\;\; i=1,2,
\label{Lax1}
\end{equation}
\noindent for the boson operators $\hat{a}_i^{\dagger}$, $\hat{a}_i$, and $\hat{N}_i$ and with $\lambda_i\gamma_i = 1$. The parameters $\alpha_i$ and $\beta_i$ are arbitrary. These operators obey the canonical boson commutation rules 

\begin{equation}
 [\hat{a}_i,\hat{a}_j]=[\hat{a}_i^{\dagger},\hat{a}_j^{\dagger}]=0,\qquad
 [\hat{a}_i,\hat{a}_j^{\dagger}] = \delta_{ij}\hat{I}, 
 \label{cr1}
\end{equation}

\begin{equation}
 [\hat{N}_i,\hat{a}_j] = -\delta_{ij} \hat{a}_j, \qquad
 [\hat{N}_i,\hat{a}_j^{\dagger}] = +\delta_{ij} \hat{a}_j^{\dagger}.
 \label{cr2}
\end{equation}
\noindent The $\hat{I}$-operator is the identity operator.

Using the co-multiplication property of the Lax operators (\ref{Lax1}) we get the following realization for the monodromy matrix,
\begin{equation}
\pi(\hat{T}(u)) = \hat{L}_{1}(u + \omega_1) \hat{L}_{2}(u - \omega_2),
\end{equation}
\noindent whose entries are,

\begin{eqnarray}
\pi(\hat{A}(u)) &=& \lambda_1\lambda_2(u + \omega_1)(u - \omega_2)\hat{I} + \lambda_1\lambda_2(u + \omega_1)\eta\hat{N}_2 \nonumber\\ 
&+& \lambda_1\lambda_2(u - \omega_2)\eta\hat{N}_1 + \lambda_1\lambda_2\eta^2\hat{N}_1\hat{N}_2 + \beta_2\alpha_1\hat{a}_2^{\dagger}\hat{a}_1, \label{A} \\
\pi(\hat{B}(u)) &=& \lambda_1\alpha_2[(u+\omega_1)\hat{I} + \eta \hat{N}_1]\hat{a}_2 + \alpha_1\alpha_2\beta_2\gamma_2\eta^{-1}\hat{a}_1, \label{B}\\
\pi(\hat{C}(u)) &=&  \beta_1\lambda_2[(u-\omega_2)\hat{I} +\eta \hat{N}_2]\hat{a}^\dagger_1 + \alpha_1\beta_1\beta_2\gamma_1\eta^{-1}\hat{a}^{\dagger}_2, \label{C} \\
\pi(\hat{D}(u)) &=& \beta_1\alpha_2\hat{a}_1^{\dagger}\hat{a}_2 + \alpha_1\alpha_2\beta_1\beta_2\gamma_1\gamma_2\eta^{-2} \hat{I}. \label{D}
\label{realiz1}
\end{eqnarray}

\noindent Hereafter we will use the same symbol for the operators and its respective realization, so we define $\pi(\hat{O}(u))\equiv \hat{O}(u)$ for any operator in the entries of the monodromy matrix (\ref{monod}).

The transfer matrix $\hat{t}(u)$ is,
\begin{eqnarray}
\hat{t}(u) &=&  \lambda_1\lambda_2(u + \omega_1)(u - \omega_2)\hat{I} + \lambda_1\lambda_2(u + \omega_1)\eta\hat{N}_2 \nonumber\\ 
&+& \lambda_1\lambda_2(u - \omega_2)\eta\hat{N}_1 + \lambda_1\lambda_2\eta^2\hat{N}_1\hat{N}_2 + \beta_2\alpha_1\hat{a}_2^{\dagger}\hat{a}_1  \nonumber \\ 
 &+& \beta_1\alpha_2\hat{a}_1^{\dagger}\hat{a}_2 + \alpha_1\alpha_2\beta_1\beta_2\gamma_1\gamma_2\eta^{-2} \hat{I}.
\label{tmr}
\end{eqnarray}
We can write the transfer matrix (\ref{tmr}) using (\ref{poli-u})
\begin{eqnarray}
\hat{t}(u) &=&  \hat{\mathcal{C}}_0 + \hat{\mathcal{C}}_1 u + \hat{\mathcal{C}}_2 u^2,
\label{tmr2}
\end{eqnarray}
\noindent with the conserved quantities

\begin{eqnarray}
\hat{\mathcal{C}}_0 &=&  \lambda_1\lambda_2(\omega_1\hat{N}_2 - \omega_2\hat{N}_1)\eta + \lambda_1\lambda_2\eta^2\hat{N}_1\hat{N}_2 \nonumber \\ 
&+& \beta_2\alpha_1\hat{a}_2^{\dagger}\hat{a}_1  + \beta_1\alpha_2\hat{a}_1^{\dagger}\hat{a}_2 + (\alpha_1\alpha_2\beta_1\beta_2\gamma_1\gamma_2\eta^{-2} - \lambda_1\lambda_2\omega_1\omega_2)\hat{I},  \\
\hat{\mathcal{C}}_1 &=&  \lambda_1\lambda_2[(\omega_1 - \omega_2)\hat{I} + \eta\hat{N}],  \\
\hat{\mathcal{C}}_2 &=&  \lambda_1\lambda_2\hat{I}.
\end{eqnarray}

We can rewrite the Hamiltonian \ref{ham} using these conserved quantities,

\begin{equation}
\hat{H} = \xi_0\hat{\mathcal{C}}_0 + \xi_1\hat{\mathcal{C}}_1^2 + \xi_2\hat{\mathcal{C}}_2,
\label{hamdecomp}
\end{equation}
\noindent with the following identification for the parameters,
\begin{eqnarray}
\xi_2 &=&  - \xi_0(\alpha_1\alpha_2\beta_1\beta_2\gamma_1^2\gamma_2^2\eta^{-2} - \omega_1\omega_2) \nonumber \\ 
      &-& \xi_1 (\omega_1 - \omega_2)^2\lambda_1\lambda_2, \\
K_{11} &=& K_{22} =  \xi_1\lambda_1^2\lambda_2^2\eta^2,  \\ 
K_{12} &=& K_{21} = (\xi_0 + 2\xi_1\lambda_1\lambda_2)\lambda_1\lambda_2\eta^2, \\
\mu_1 - U_1  &=&  [2\xi_1\lambda_1\lambda_2\omega_1 - (\xi_0 + 2\xi_1\lambda_1\lambda_2)\omega_2]\lambda_1\lambda_2\eta, \\
\mu_2 - U_2  &=&  [(\xi_0 + 2\xi_1\lambda_1\lambda_2)\omega_1 - 2\xi_1\lambda_1\lambda_2\omega_2]\lambda_1\lambda_2\eta, \\
\Omega_{12} &=& -\xi_0\beta_1\alpha_2, \\ 
\Omega_{21} &=&  -\xi_0\beta_2\alpha_1, 
\end{eqnarray}
\noindent with $\xi_i \neq 0,\; i =0,1,2.$

Now it is straightforward to check that the Hamiltonians (\ref{ham}) and (\ref{hamdecomp}) are  related to the transfer matrix  $\hat{t}(u)$ (\ref{tmr}) through
\begin{equation}
\hat{H} = \xi_0\hat{t}(u) + \xi_1\hat{\mathcal{C}}_1^2 - \xi_0\hat{\mathcal{C}}_1 u  - (\xi_0 u^2 - \xi_2)\hat{\mathcal{C}}_2,
\label{hamtu}
\end{equation}
\noindent  and from \ref{hamdecomp} or \ref{hamtu} that $[\hat{H},\hat{t}(u)]=0$. Notice that the spectral parameter $u$ appearing in the Hamiltonian (\ref{hamtu}) is canceled. The Hamiltonian parameters in (\ref{ham}) or (\ref{hamtu}) are real numbers. The transfer matrix parameters in (\ref{tmr}) can be complex numbers, but in this case the transfer matrix is not Hermitian. We will only consider the Hermitian case.

We can apply the algebraic Bethe ansatz method, using the  Fock vacuum as the pseudo-vacuum $|0\rangle = |0\rangle_1\otimes|0\rangle_2$, to find the BAE
\begin{equation}
\frac{\lambda_1\lambda_2[v_i^2 + (\omega_1 - \omega_2)v_i - \omega_1\omega_2]}{\alpha_1\alpha_2\beta_1\beta_2\gamma_1\gamma_2\eta^{-2}} =
\prod ^N_{j \neq i}\frac {v_i -v_j - \eta}{v_i - v_j + \eta}, \;\;\;\;\;i,j = 1,\ldots,N,
\label{becbae} 
\end{equation}
\noindent and the energies of the Hamiltonian

\begin{eqnarray}
 E(\{v_i\}) &=& \xi_0\lambda_1\lambda_2[u^2 + (\omega_1 - \omega_2)u -\omega_1\omega_2]\prod_{i=1}^N\left(1+\frac{\eta}{v_i-u}\right) \nonumber \\ 
 &+& \xi_1\lambda^2_1\lambda^2_2(\omega_1 - \omega_2 + \eta N)^2 - \xi_0\lambda_1\lambda_2(\omega_1 - \omega_2 + \eta N) u  \nonumber \\ 
 &-& \xi_0 \lambda_1\lambda_2 u^2 + \xi_2 \lambda_1\lambda_2 +  \xi_0\alpha_1\alpha_2\beta_1\beta_2\gamma_1\gamma_2\eta^{-2}\prod_{i=1}^N\left(1-\frac{\eta}{v_i-u}\right). \nonumber \\  
\label{becnrg}
\end{eqnarray}
Fortunately this expression is a function of the spectral parameter $u$, which can be chosen arbitrarily.  For asymmetric tunnelling, $\Omega_{12} \neq \Omega_{21}$, we can consider $\alpha_1\beta_2 = \eta$ and $\beta_1\alpha_2 = \kappa\eta$ and in the limit of no interaction, $K_{ij} \rightarrow 0$ with $\eta \ll 1$, we can write the Bethe ansatz equation (\ref{becbae}) as 
\begin{equation}
\sum_{i=1}^N \left[ v_i + \frac{1}{2}(\omega_1 - \omega_2)\right]^2 = R_N^2.
\label{becbae2}
\end{equation}
\noindent The Eq. (\ref{becbae2}) is the equation of a complex manifold in $\mathbb{C}^N$ with

\begin{equation}
R_N = \sqrt{\left[ \frac{1}{4}(\omega_1 - \omega_2)^2 + \frac{(\kappa + \lambda_1^2\lambda_2^2\omega_1\omega_2)}{\lambda_1^2\lambda_2^2}\right]N}.
\end{equation}

\noindent If all Bethe roots  $\{v_i\}$ are real numbers, in $\mathbb{R}^N$ the surface is a $S^{N-1}$ sphere  with radii $R_N$ and center in

\begin{equation}
v_i = - \frac{1}{2}(\omega_1 - \omega_2), \;\;\;\;\; \forall i = 1, \ldots, N.
\end{equation}

For $u=0$ and $K_{ij} \rightarrow 0$ we can write the eigenvalues as
\begin{eqnarray}
 E(\{v_i\}) &=& \xi_1\lambda_1^2\lambda_2^2(\omega_1 - \omega_2 + \eta N)^2 + \xi_2\lambda_1\lambda_2 + \xi_0(\kappa\gamma_1\gamma_2 - \lambda_1\lambda_2\omega_1\omega_2) \nonumber \\
 &-& \xi_0(\lambda_1\lambda_2\omega_1\omega_2 +  \kappa\gamma_1\gamma_2)\eta\sum_{i=1}^N\frac{1}{v_i}.
\label{becnrgsum}
\end{eqnarray}

When we consider symmetric tunnelling we just put $\kappa = 1$ and $\omega_1 = -\omega_2$ to get

\begin{eqnarray}
 E(\{v_i\}) &=& \xi_1\lambda_1^2\lambda_2^2(2\omega_1 + \eta N)^2 + \xi_2\lambda_1\lambda_2 + \xi_0(\gamma_1\gamma_2 + \lambda_1\lambda_2\omega_1^2) \nonumber \\
 &-& \xi_0(\gamma_1\gamma_2 - \lambda_1\lambda_2\omega_1^2)\eta\sum_{i=1}^N\frac{1}{v_i}.
\label{becnrgsum2}
\end{eqnarray}

\section{Summary}

 I have introduced a new parametrization of a bosonic Lax operator and explicitly calculated the spectrum of a generalized two-sites Bose-Hubbard model with asymmetric tunnelling by the algebraic Bethe  ansatz method using the $gl(2)$-invariant $R$-matrix and showed that in the no interaction limit the Bethe ansatz equations can be written as a $S^{N-1}$ sphere in $\mathbb{R}^N$, where $N$ is the total number of atoms in the condensate.

\section*{Acknowledgments}

The author acknowledge CAPES/FAPERJ (Coordena\c{c}\~ao de Aperfei\c{c}oamento de Pessoal de N\'{\i}vel Superior/Funda\c{c}\~ao de Amparo \`a Pesquisa do Estado do Rio de Janeiro) for the financial support.


\begin{thebibliography}{10}

\bibitem{ak} J. R. Anglin  and  W. Ketterle, \textit{Nature} \textbf{416} (2002) 211.

\bibitem{anderson} M. H. Anderson, J. R. Ensher,  M. R. Mathews, C. E. Wieman and E. A. Cornell, \textit{Science}  {\bf 269} (1995) 198.

\bibitem{wwcch} J. Williams, R. Walser, J. Cooper, E. A. Cornell and M. Holland, \textit{Phys. Rev. A} \textbf{61} (2000) 0336123.
\bibitem{bose} S. N. Bose, \textit{Z. Phys.}  \textbf{26} (1924) 178.

\bibitem{eins}  A. Einstein, \textit{Phys. Math. K1}  \textbf{22} (1924) 261.
\bibitem{Dalfovo} F. Dalfovo,  S. Giorgini,  L. P. Pitaevskii and  S. Stringari,  \textit{Rev. Mod. Phys.}  \textbf{71} (1999) 463.

\bibitem{l01} A. J. Leggett, \textit{Rev. Mod. Phys.}  \textbf{73} (2001) 307.

\bibitem{cw} E. A. Cornell and C. E. Wieman, \textit{Rev. Mod. Phys.} \textbf{74} (2002) 875.

\bibitem{Donley} E. A. Donley, N. R. Claussen,  S. T. Thompson and C. E. Wieman, \textit{Nature} \textbf{417} (2002) 529.

\bibitem{Bloch} I. Bloch,  J. Dalibard and W. Zwerger, \textit{Rev. Mod. Phys.}  \textbf{80} (2008) 875.

\bibitem{Carusotto} I. Carusotto and  C. Ciuti, \textit{Rev. Mod. Phys.}  \textbf{85} (2013) 299.
\bibitem{GSantos11a} G. Santos,  A. Foerster,  I. Roditi, Z. V. T. Santos and A. P. Tonel, \textit{J. Phys. A: Math. Theor.} \textbf{41} (2008) 295003 (9pp).
 
\bibitem{GSantos11b} G. Santos, \textit{J. Phys. A: Math. Theor.} \textbf{44} (2011) 345003.

\bibitem{GSantos13}  G. Santos,  A. Foerster and I. Roditi, \textit{J. Phys. A: Math. Theor.} \textbf{46} (2013) 265206 (12pp).

\bibitem{GSantos06a} G. Santos, A. Tonel,  A. Foerster and J. Links, \textit{Phys. Rev. A} \textbf{73} (2006) 023609.

\bibitem{GSantos09} A. P. Tonel, C. C. N. Kuhn,  G. Santos, A. Foerster, I. Roditi and Z. V. T. Santos,  \textit{Phys. Rev. A} \textbf{79} (2009) 013624.

\bibitem{GSantos10} G. Santos, A. Foerster, J. Links,  E. Mattei and S. R. Dahmen, \textit{Phys. Rev. A} \textbf{81} (2010)  063621.

\bibitem{albiez} M. Albiez, R. Gati,  J. F\"olling, S. Hunsmann,  M.  
Cristiani  and M. K. Oberthaler, \textit{Phys. Rev. Lett.} \textbf{95} (2005) 010402.

\bibitem{milb} G. J. Milburn, J. Corney,  E. M.  Wright and D. F. Walls, \textit{Phys. Rev. A}  \textbf{55} (1997) 4318.

\bibitem{hines2} A. P. Hines, R.  H. McKenzie and G. J. Milburn, \textit{Phys. Rev. A} \textbf{67} (2003) 013609.

\bibitem{ours} A. P. Tonel, J. Links and A. Foerster, \textit{J. Phys. A: Math. Gen.} \textbf{38} (2005) 6879.

\bibitem{our} A. P. Tonel, J. Links and  A. Foerster, \textit{J. Phys. A: Math. Gen.} \textbf{38} (2005) 1235.

\bibitem{hines} A. P. Hines,  R. H. McKenzie  and G. J. Milburn, \textit{Phys. Rev. A} \textbf{71} (2005) 042303.

\bibitem{GSantos06b}  J. Links, A. Foerster, A. P. Tonel and G. Santos, \textit{Ann. Henri Poincar\'e}  \textbf{7}  (2006) 1591.

\bibitem{jlreview} J. Links, H.-Q. Zhou,  R. H. McKenzie  and M. D. Gould, \textit{J. Phys. A: Math. Gen.} \textbf{36} (2003) R63.

\bibitem{sarma} Qi Zhou and Das Sarma, \textit{Phys. Rev. A} \textbf{82} (2010) 041601(R). 

\bibitem{Dounas} D. R. Dounas-Frazer, A. M. Hermundstad and  L. D. Carr,  \textit{Phys. Rev. Lett.} \textbf{99} (2007) 200402. 

\bibitem{Shin} Y. Shin, M. Saba, A. Schirotzek, T. A. Pasquini, A. E. Leanhardt, D. E. Pritchard and W. Ketterle, 
\textit{Phys. Rev. Lett.} \textbf{92} (2004) 150401. 

\bibitem{Seaman} B. T. Seaman, M. Kr\"amer, D. Z. Anderson and M. J. Holland, \textit{Phys. Rev. A} \textbf{75} (2007)  023615.

\bibitem{Labouvie} R. Labouvie, B. Santra, S. Heun, S. Wimberger and H. Ott, \textit{Phys. Rev. Lett.} \textbf{115} (2015) 050601.


\bibitem{l02} I. Zapata, F. Sols and A. J. Leggett, \textit{Phys. Rev. A} \textbf{57} (1998) R28.
\bibitem{jlletter} J. Links  and H.-Q. Zhou, \textit{Lett. Math. Phys.} \textbf{60} (2002) 275.

\bibitem{Angela} A. Foerster, J. Links and H.-Q. Zhou,  
in {\it Classical and quantum nonlinear integrable systems: theory and applications}, edited
by A. Kundu (Institute of Physics Publishing, Bristol and Philadelphia, 2003)
pp 208--233. 

\bibitem{jlsigma} J. Links  and K. E. Hibberd, \textit{SIGMA} \textbf{2} (2006) 095 (8pp).

\bibitem{Angela2} D. Rubeni, A. Foerster,  E. Mattei and  I. Roditi, \textit{Nuc. Phys. B} \textbf{856} (2012) 698.

\bibitem{Bethe-states} G. Santos, C. Ahn, A. Foerster  and I. Roditi, \textit{Phys. Lett. B} \textbf{746} (2015) 186.

\end{thebibliography}
\end{document}